\pdfoutput=1

\documentclass{article}
\usepackage[pass]{geometry}
\usepackage{times}
\usepackage{fullpage}
\usepackage{paralist}
\usepackage{url}
\usepackage{dcolumn}
\usepackage{numprint}

\usepackage{booktabs, threeparttable, tabularx, multirow}
\usepackage{graphicx, subcaption, xcolor}
\usepackage{amsmath, amssymb, amsfonts, bm}
\usepackage{algorithm, algpseudocode}
\usepackage[authoryear]{natbib}
\usepackage{underscore}

\usepackage[utf8x]{inputenc}
\usepackage[T1]{fontenc}

\usepackage[colorinlistoftodos]{todonotes}
\usepackage{hyperref}

\newcommand{\TeamName}{\texttt{IlpsUvA}}
\newcommand{\BOIRRun}{\texttt{IlpsUvABoir}}
\newcommand{\NVSMRun}{\texttt{IlpsUvANvsm}}
\newcommand{\NVSMQLMRun}{\texttt{IlpsUvAQlmNvsm}}

\newcommand{\bo}{Bayesian Optimization }

\title{ILPS at TREC 2017 Common Core Track}
\author{Christophe Van Gysel, Dan Li, Evangelos Kanoulas\footnote{Any permutation of this ordered list is also valid.}\\
cvangysel@uva.nl, d.li@uva.nl, e.kanoulas@uva.nl\\
Informatics Institute\\
University of Amsterdam\\
The Netherlands}

\date{}

\begin{document}
\maketitle

\begin{abstract}
The TREC 2017 Common Core Track aimed at gathering a diverse set of participating runs and building a new test collection using advanced pooling methods.
In this paper, we describe the participation of the \TeamName{} team at the TREC 2017 Common Core Track. We submitted runs created using two methods to the track:
\begin{inparaenum}[(1)]
	\item BOIR uses Bayesian optimization to automatically optimize retrieval model hyperparameters. 
	\item NVSM is a latent vector space model where representations of documents and query terms are learned from scratch in an unsupervised manner.
\end{inparaenum}

We find that BOIR is able to optimize hyperparameters as to find a system that performs competitively amongst track participants. NVSM provides rankings that are diverse, as it was amongst the top automated unsupervised runs that provided the most unique relevant documents.

\end{abstract}

\renewcommand{\cite}{\citep}

\section{Introduction}

TREC 2017 Core Track aims to bring the information retrieval community back into a traditional ad-hoc search task. The primary goal of itself is to build new test collections using recently created documents using new test collection construction methodology based on a diverse set of participating runs.
In this work, we applied two methods on the ad-hoc task: a Bayesian Optimization intensified lexical method (BOIR), and a latent vector space method, named neural vector space model (NVSM).

BOIR uses Bayesian Optimization method to automatically optimize configurations of Retrieval system. In this work, we take Indri as our experiment platform. Indri is a search engine pipeline consisting of many components such as indexing module, retrieval module, and pseudo-relevance feedback module. It has a big configuration (or hyperparameter) space. We use Bayesian Optimization, a sequential decision making method which suggests the next most promising configuration to be tested on the basis of the retrieval effectiveness of hyperparameters that have been examined so far, to jointly search and optimize over the hyperparameter space. As the retrieval models in Indri are TF-IDF, BM25, Language model, we submitted BOIR run - \BOIRRun{} - as a lexical run.

NVSM, on the other hand, is a method that learns representations of documents in an unsupervised manner for news article retrieval. In the NVSM paradigm, we learn low-dimensional representations of words and documents from scratch using gradient descent and rank documents according to their similarity with query representations that are composed from word representations. Contrast to BOIR, NVSM overcomes the vocabulary mismatch between query and documents and provide  semantic matching. We therefore also submitted two runs generated by NVSM: \NVSMRun{} and \NVSMQLMRun{}.

This paper is organized as follows: Section~\ref{method} describes our approach to this task. Section~\ref{result} discusses the results obtained from applying this approach to the test datasets. Finally, Section~\ref{conclusion} presents the conclusions and potential future work.

\section{Methodology}
\label{method}
\subsection{Neural Models for IR}
\textbf{Neural Vectors Spaces Models for Unsupervised Retrieval}. We constructed the vector space models from \citet{VanGysel2017nvsm,van2016learning} on the New York Times corpus. The rankings of the \NVSMRun{} run were obtained using the methodology as outlined in the \emph{Unsupervised Deployment} section of \citet{VanGysel2017nvsm}. To construct the rankings of the \NVSMQLMRun{} run, we took the scores of the \NVSMRun{} rankings and combined them with the log-probability retrieval scores of the Query-Likelihood Model \citep{Zhai2004smoothing} with Dirichlet smoothing ($\mu = 1000$, i.e., Indri's default value). The combination was obtained by summing the per-query standarized scores of the QLM and the ensemble of NVSM. An open-source implementation of NVSM is available at \url{https://github.com/cvangysel/cuNVSM}.

\subsection{Bayesian Optimization for IR}

\begin{figure}[h]
\centering
\includegraphics[scale=0.7]{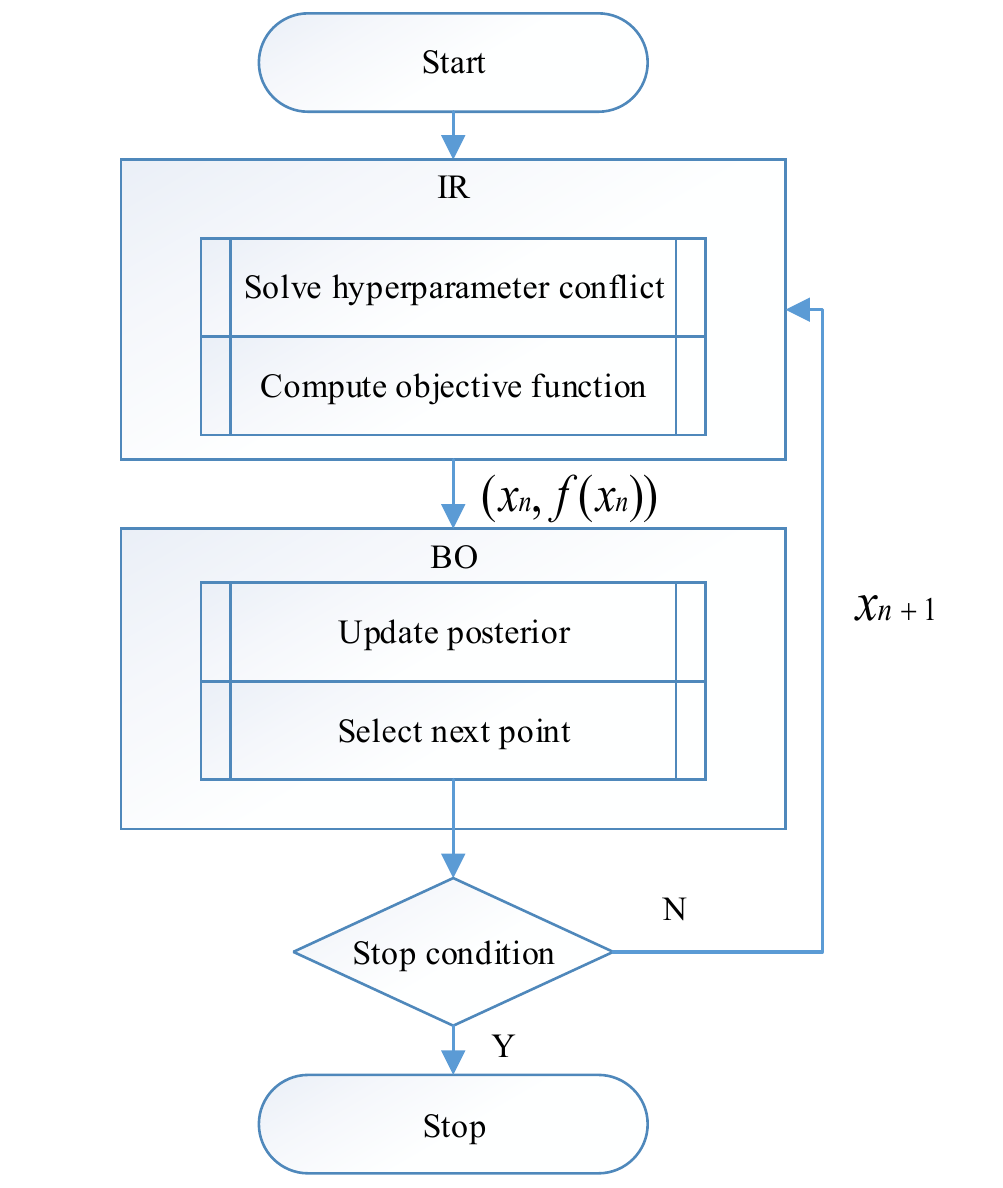}
\caption{Hyperparameter optimization architecture.}
\label{fig:architecture}
\end{figure}

\textbf{Hyperparameter Optimization Process}. The \bo framework provides a mechanism to sequentially search for the global optimum $x$ of an objective function $f(x):\mathbb{X}\to \mathbb{R}$. There are two key components in \bo~\cite{shahriari2016taking}. The first is a probabilistic \textit{surrogate model} used to predict the objective function value $y$ given a point $x$. For every $x$, there is a random variable $y$, whose distribution $p(y|x)$ is given by the surrogate model. The predictive distribution of $y$ can then be used to construct an \textit{acquisition function}. An acquisition function is a policy for selecting the sequence of points $\{x_1, x_2, ..., x_i, ...\}$, i.e. a mechanism to select the next configuration $x_{n+1}$ to test given $D_{1:n}:=\{(x_1, y_1), (x_2, y_2), ... ,(x_n, y_n)\}$. As the acquisition function x is  usually a closed-form expression of hyperparameters it is easier to be optimized than the original objective function. The second component is the \textit{objective function} itself, a function of the target model requiring hyperparameter optimization. 

Following the \bo framework for optimizing retrieval systems, proposed by~\cite{Dan2017}, we have two major modules in our algorithmic pipeline, the IR module and the BO module (see Figure~\ref{fig:architecture}). The IR module tackles the conditional hyperparameters, and computes the objective function value $y_n$, given a hyperparameter configuration $x_n$. The BO module adds $(x_n,y_n)$ into the sample set, updates the posterior distribution of the surrogate models, selects the next hyperparameter configuration $x_{n+1}$, and passes it back to IR module. The process stops when the stop condition is satisfied, such as the computation budget is run out.

\textbf{Implementation of Retrieval System and Bayesian Optimization}. We use Pyndri~\cite{van2017pyndri}, a Python Interface to the \textit{Indri} Search Engine ~\cite{Strohman05indri} \footnote{https://www.lemurproject.org/indri.php}, as the IR module in our pipeline, which is mainly decomposed to indexing, retrieval, and pseudo-relevance feedback modules. All the three modules are considered in our optimization experiments. The objective function can be any retrieval effectiveness measure. In this work we optimize for the Mean Average Precision (MAP) and use \textit{trec_eval}\footnote{\url{http://trec.nist.gov/trec_eval/}} for the computations. We use \textit{Pybo}~\cite{hoffman2014modular} in our experiments, a Python package for \bo. We set Gaussian process with Squared Exponential covariance function as the surrogate model, and Expected Improvement~\cite{dixon1978towards} as the acquisition function. The hyperparameter optimization precess was conducted on the Rubost04 topics and the corresponding document collection of Volume 4 \& 5 (minus Congress Record).

\textbf{Search Space of Hyperparameters}. When configuring \textit{Indri}, there are two choices to make when indexing documents: the stopword list and the stemmer, and three retrieval models to choose: TF-IDF with BM25 term weighting, Okapi BM25, and Language Models. 
\textit{Indri} also supports pseudo-relevance feedback models. Each of these choices contains subsequent parameters to be decided.

\begin{table}[ht]
\centering
\caption{Hyperparameters and their search ranges in Indri}
\label{tbl:params}

\centering
\begin{tabular}{lll}
\toprule
Hyperparamter & Type & Values \\ \hline
$Stopper$ & Boolean & \{True, False\} \\
$Stemmer$ & Boolean & \{True, False\} \\ \hline
$rm$ & Integer & \{TF-IDF, BM25, LM-JM, LM-DIR, LM-TS\} \\
$k1$ & Real value & {[}1,2{]} \\
$b$ & Real value & {[}0,1{]} \\
$k1$ & Real value & {[}1,10{]} \\
$k3$ & Real value & {[}1,10{]} \\
$b$ & Real value & {[}0,1{]} \\
$\lambda_{doc}$ & Real value & {[}0,1{]} \\
$\lambda_{col}$ & Real value & {[}0,1{]} \\
$\mu_{dir}$ & Real value & {[}0,3000{]} \\
$\mu_{ts}$ & Real value & {[}0,3000{]} \\
$\lambda_{ts}$ & Real value & {[}0,1{]} \\ \hline
$prf$ & Boolean & \{True, False\} \\
$fbDocs$ & Integer & {[}1,50{]} \\
$fbTerms$ & Integer & {[}1,50{]} \\
$fbMu$ & Real value & {[}0,3000{]} \\
$fbOrigWeight$ & Real value & {[}0,1{]} \\
\bottomrule
\end{tabular}

\end{table}

In total, we have a conditional hyperparameter space of 18 dimensions (see Table \ref{tbl:params}).

\section{Results}
\label{result}
The \TeamName{} team submitted 3 automatic runs to TREC Core Track. \BOIRRun{} is an automated run that made use of the existing Robust04 judgments. The two automated remaining runs, \NVSMRun{} and \NVSMQLMRun{}, were constructed without any prior knowledge of the topics.

TREC Core Track provides two types of relevance assessments: judgments provided by NIST assessors and judgments obtained through crowd sourcing. In this paper, we report the result evaluated on the 50 topics judged by the NIST assessors.

\subsection{Test Dataset}
TREC Core track uses The New York Times corpus \footnote{https://catalog.ldc.upenn.edu/ldc2008t19} as the document collection, and the TREC Robust Track topics as its topics. Most of the topics have remained the same but some have been revised to reflect the time past. 
There are two sets of topics: topics judged by NIST and topics to be judged by crowd sourcing. Submission should either be either on the 50 topics to be judged by NIST or on all 250 topics to be judged by crowd workers.  The 3 runs submitted by \TeamName{} include all the 250 topics.

\subsection{\TeamName{} runs}

Table \ref{tbl:overall-ret} compares the overall performance of ILPS within the TREC Core Track. The entries in the table listed as \emph{best}, \emph{median} and \emph{worst} correspond to the average of the respective aggregate across all topics (i.e., a virtual system). That is, the \emph{best automated} run is the average (across topics) of the maximum AP (across participating runs). It is important to note that the aggregate measures of the automated runs provided by the organizers were computed from runs that used existing relevance judgments. \BOIRRun{} performs worse than the best automated run. To little surprise, \BOIRRun{} performs better than the worst automated run. \BOIRRun{} also performs better than the median automated run in terms of AP and NDCG. When optimizing for the Robust04 topics and relevance judgments, the Bayesian optimization algorithm chose the following hyperparameters: the QLM \citep{Zhai2004smoothing} with Dirichlet smoothing ($\mu=721$) and pseudo-relevance feedback for query expansion. The result indicates that the classical lexical models are quite strong baselines. 

The \NVSMRun{} and \NVSMQLMRun{} perform worse than the virtual automated \emph{median} system. This is not surprising, as the \emph{median} retrieval effectiveness was computed from automated runs that additionally used existing relevance judgments. However, it is important to note that the \NVSMRun{} and \NVSMQLMRun{} were amongst the top-3 automated systems (using no existing relevance judgments) that contributed the most unique relevant documents to the test collection. This is due to the fact that NVSM provides semantic matching and, consequently, retrieves relevant documents that were previously not retrieved by methods that perform retrieval using exact term matching.

\begin{figure}
\centering
\includegraphics[scale=.5]{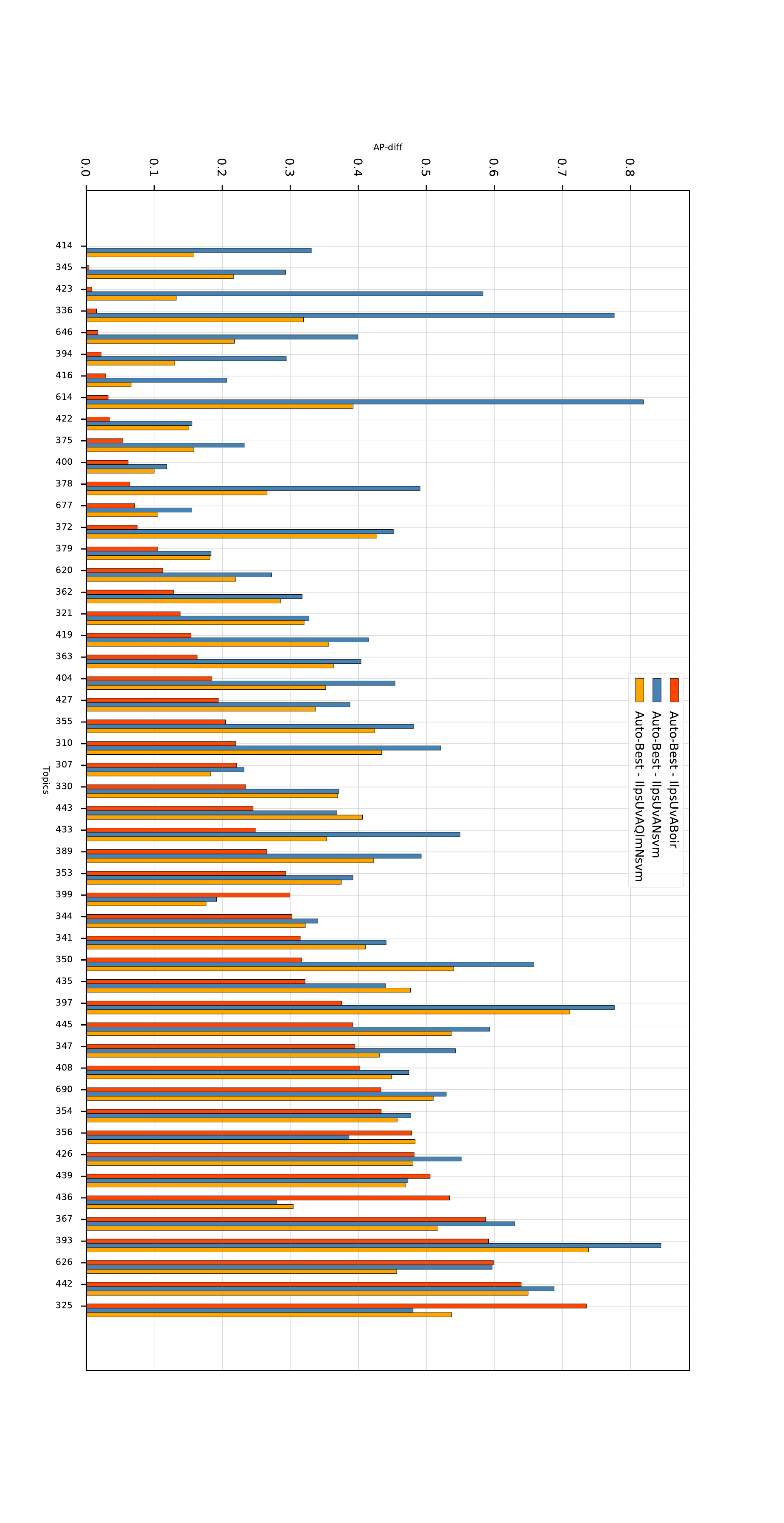}
\caption{Performance difference between \TeamName{} and the Auto-Best run in TREC Core.}
\label{fig:boirrettopics}
\end{figure}

We also plot the result on each of the 50 topics (see Figure \ref{fig:boirrettopics}). By each topic, we calculate the difference of AP scores between \TeamName{} runs and the best run in TREC Core in order to discover how the lexical method and the semantic method differentiate with each other. The result shows that they are complementary to some extent. For example, on topic 325 (Cult Lifestyles), 336 (Black Bear Attacks), and 345 (Overseas Tobacco Sales). This reminds us that the future work may lie in combing the two kinds of philosophy and take their respective advantages. 

\begin{table}[ht]
\centering
\caption{The overall performance of \TeamName{} in TREC Core Track. \BOIRRun{} is an automated run that used existing relevance judgments, whereas \NVSMRun{} and \NVSMQLMRun{} are automated runs that used no prior information. The \texttt{Auto-*} and \texttt{Manu-*} runs were provided by the track's organizers and contain runs that made use of existing judgments.}
\label{tbl:overall-ret}
\begin{tabular}{ llll}
\toprule
Run & AP & NDCG & P@10 \\ \hline
\BOIRRun{} & \nprounddigits{3}\npdecimalsign{.}\npthousandsep{.}\numprint{0.2861} & \nprounddigits{3}\npdecimalsign{.}\npthousandsep{.}\numprint{0.5151} & \nprounddigits{3}\npdecimalsign{.}\npthousandsep{.}\numprint{0.5700} \\
\NVSMRun{} & \nprounddigits{3}\npdecimalsign{.}\npthousandsep{.}\numprint{0.1260} & \nprounddigits{3}\npdecimalsign{.}\npthousandsep{.}\numprint{0.3325}  & \nprounddigits{3}\npdecimalsign{.}\npthousandsep{.}\numprint{0.3220} \\
\NVSMQLMRun{} & \nprounddigits{3}\npdecimalsign{.}\npthousandsep{.}\numprint{0.1715} & \nprounddigits{3}\npdecimalsign{.}\npthousandsep{.}\numprint{0.4122} & \nprounddigits{3}\npdecimalsign{.}\npthousandsep{.}\numprint{0.4180} \\ \hline
\texttt{Auto-Best} & \nprounddigits{3}\npdecimalsign{.}\npthousandsep{.}\numprint{0.537678} & \nprounddigits{3}\npdecimalsign{.}\npthousandsep{.}\numprint{0.769866} & \nprounddigits{3}\npdecimalsign{.}\npthousandsep{.}\numprint{0.916} \\
\texttt{Auto-Median} & \nprounddigits{3}\npdecimalsign{.}\npthousandsep{.}\numprint{0.22803} & \nprounddigits{3}\npdecimalsign{.}\npthousandsep{.}\numprint{0.47865} & \nprounddigits{3}\npdecimalsign{.}\npthousandsep{.}\numprint{0.548} \\
\texttt{Auto-Worst} & \nprounddigits{3}\npdecimalsign{.}\npthousandsep{.}\numprint{0.006036} & \nprounddigits{3}\npdecimalsign{.}\npthousandsep{.}\numprint{0.048072} & \nprounddigits{3}\npdecimalsign{.}\npthousandsep{.}\numprint{0.002} \\ \hline
\texttt{Manu-Best} & \nprounddigits{3}\npdecimalsign{.}\npthousandsep{.}\numprint{0.542618} & \nprounddigits{3}\npdecimalsign{.}\npthousandsep{.}\numprint{0.769536} & \nprounddigits{3}\npdecimalsign{.}\npthousandsep{.}\numprint{0.922} \\
\texttt{Manu-Median} & \nprounddigits{3}\npdecimalsign{.}\npthousandsep{.}\numprint{0.378618} & \nprounddigits{3}\npdecimalsign{.}\npthousandsep{.}\numprint{0.637954} & \nprounddigits{3}\npdecimalsign{.}\npthousandsep{.}\numprint{0.672} \\
\texttt{Manu-Worst} & \nprounddigits{3}\npdecimalsign{.}\npthousandsep{.}\numprint{0.165292} & \nprounddigits{3}\npdecimalsign{.}\npthousandsep{.}\numprint{0.399122} & \nprounddigits{3}\npdecimalsign{.}\npthousandsep{.}\numprint{0.282} \\

\bottomrule
\end{tabular}
\end{table}
\section{Conclusion}
\label{conclusion}

In this paper, we described the participation of the \TeamName{} team at TREC 2017 Core Track. We applied two methods on the ad-hoc task: a Bayesian Optimization intensified lexical method (BOILM), and a latent vector space method, NVSM. Overall, \TeamName{} contributed 3 runs to the track: \BOIRRun{}, \NVSMRun{} and \NVSMQLMRun{}.

\bibliographystyle{abbrvnatnourl}
\bibliography{sample}

\end{document}